\documentclass[12pt]{iopart}

\newcommand{\ave}[1]{\ensuremath{\left\langle#1\right\rangle}}
\newcommand{\aves}[1]{\ensuremath{\langle#1\rangle}}
\usepackage{iopams}
\usepackage{graphicx}
\usepackage{color}
\usepackage{cite}
\usepackage{bm}
\begin{document}

\title[Singular behaviour of time-averaged stress fluctuations on surfaces]{Singular behaviour of time-averaged stress fluctuations on surfaces}

\author{Masato Itami$^1$ and Shin-ichi Sasa$^2$}
\address{$^1$ Fukui Institute for Fundamental Chemistry, Kyoto University, Kyoto 606-8103, \hspace*{2ex}Japan}
\address{$^2$ Department of Physics, Kyoto University, Kyoto 606-8502, Japan}
\eads{\mailto{itami@fukui.kyoto-u.ac.jp} and \mailto{sasa@scphys.kyoto-u.ac.jp}}

\begin{abstract}
We provide a method for calculating time-averaged stress fluctuations on surfaces in a viscous incompressible fluid at equilibrium.
We assume that (i) the time-averaged fluctuating stress is balanced in equilibrium at each position and that (ii) the time-averaged fluctuating stress obeys a Gaussian distribution on the restricted configuration space given by (i). 
Using these assumptions with the Green--Kubo formula for the viscosity, we can derive the large deviation function of the time-averaged fluctuating stress.
Then, using the saddle-point method for the large deviation function, we obtain the time-averaged surface stress fluctuations.
As an example, for a fluid between two parallel plates, we study the time-averaged shear/normal stress fluctuations per unit area on the top plate at equilibrium. 
We show that the surface shear/normal stress fluctuations are inversely proportional to the distance between the plates.
\vspace*{1ex}\\
\textbf{Keywords:} current fluctuations, fluctuating hydrodynamics, large deviation
\end{abstract}
%
%
%
%
%
\section{Introduction}

Fluctuations and correlations play significant roles in statistical physics.
In equilibrium statistical mechanics, fluctuations of thermodynamic quantities and their spatial correlations near a critical point universally characterize a continuous phase transition~\cite{Landau-LifshitzStat}.
In non-equilibrium statistical mechanics, fluctuations of time-averaged quantities and time correlations in equilibrium systems are related to responses to small perturbations or transport coefficients, following linear response theory~\cite{Groot-Mazur,Kubo-Toda-Hashitsume}.
In the last two decades, non-trivial relations that are generally valid far from equilibrium, including the fluctuation theorem~\cite{Evans-Cohen-Morriss,Gallavotti-Cohen,Kurchan,Lebowitz-Spohn,Maes,CrooksPRE1,CrooksPRE2,JarzynskiJSP,SeifertPRL} and the Jarzynski equality~\cite{JarzynskiPRL}, have been developed as a result of the time-reversal symmetry of microscopic mechanics.
These relations provide a unified view of the well-known relations, such as McLennan ensembles, the fluctuation-dissipation relation, and the Kawasaki nonlinear response relation~\cite{CrooksPRE2,Hayashi-Sasa}.
Using these techniques, and assuming a local Gibbs distribution at the initial time, the Navier--Stokes equation was derived for an isolated Hamiltonian system~\cite{Sasa}.

Fluctuations and correlations in dilute fluids have been investigated in the framework of the Boltzmann equation~\cite{Chapman-Cowling}, while those in dense fluids have been investigated in the framework of fluctuating hydrodynamics~\cite{Landau-LifshitzFluid,Schmitz}; in the latter case, Gaussian white noise satisfying the fluctuation-dissipation relation has been added to the standard hydrodynamic equations.
For example, Zwanzig~\cite{Zwanzig} calculated the time correlation of a force acting on a fixed sphere in a viscous incompressible fluid, which is related to the friction coefficient of the sphere in terms of linear response theory~\cite{Kirkwood}.
Based on Fax\'en's law~\cite{Kim-Karrila}, the time correlation of the force is related to the time correlation of the velocity fields without the sphere, which can be calculated by using fluctuating hydrodynamics.
The original form of Fax\'en's law was only applicable to stationary flow, so its use by Zwanzig was unjustified.
Bedeaux and Mazur derived a generalized form of Fax\'en's law using the method of induced forces and calculated the time correlation of a force acting on a sphere moving with a time-dependent velocity in a viscous incompressible fluid~\cite{Bedeaux-Mazur}.
These results were derived for an infinite system.
In Ref.~\cite{MonahanETAL}, for a viscous compressible fluid between two parallel plates, the time correlation of normal stresses acting on the plates was calculated by using the Green's function method.
However, since their results depend on the microscopic scale at which the continuum hydrodynamic description breaks down, an additional assumption may be necessary for a quantitative calculation.
Note that, very recently, the time correlation of shear stresses acting on plates with a finite slip length was calculated in Ref.~\cite{Nakano-Sasa} by using the same method as in Ref.~\cite{MonahanETAL}.
Even today, it remains difficult to calculate time-averaged surface stress fluctuations in a finite space.

In this paper, we propose a new approach to calculate time-averaged stress fluctuations on surfaces by using the contraction principle.
We derived Stokes' law from a microscopic point of view in our previous paper~\cite{Itami-Sasa2015}, where a large deviation function on a surface was determined from the variational principle associated with a large deviation function in the bulk.
Extending the method of \cite{Itami-Sasa2015} by considering the influence of the volume viscosity, we investigate time-averaged shear and normal stress fluctuations on surfaces in a viscous incompressible fluid at equilibrium.
This is not a simple exercise of the calculation.
We find that the fluctuations of the time-averaged stresses exhibit singular and unintuitive behaviour.
As an example, for a fluid between two parallel plates, we find that the time-averaged shear and normal stress fluctuations on the plates are inversely proportional to the distance between the plates.
Note that the anomalous system size dependence of a large deviation function for a time-averaged density in low dimensions was investigated by using the contraction principle in Ref.~\cite{Shiraishi}, where fluctuations of the time-averaged density in one dimension were proportional to the system size due to the long-time tail behaviour in low dimensions.

The remainder of this paper is organized as follows.
In section~\ref{sec:setup}, we explain the setup of our model.
In section~\ref{sec:results}, extending the method of \cite{Itami-Sasa2015}, we investigate time-averaged shear/normal stress fluctuations on surfaces.
The final section is devoted to a brief summary and some concluding remarks.

Throughout this paper, the subscripts $a,b,c$ represent indices in the Cartesian coordinates $(x,y,z)$.

\section{Setup} \label{sec:setup}

\subsection{Model}

We consider a viscous incompressible fluid between two parallel plates of length $L_{x}$ and width $L_{y}$.
The plates are oriented horizontally in the $xy$-plane and separated by a distance $L_{z}$.
We impose periodic boundary conditions in the $x$- and $y$-directions.
Let $\sigma_{ab}(\bm{r},t)$ denote the $ab$-component of the fluctuating viscous stress tensor $\bm{\sigma}$ of the fluid at position $\bm{r}=(x,y,z)$ and time $t$ in Cartesian coordinates.
Because we consider only simple fluids, we assume that the stress tensor is symmetric.
Thus, we identify $\sigma_{ab}$ with $\sigma_{ba}$.
Note that when each atom-pair interaction potential depends only on the distance between the atoms, the microscopic stress tensor is always symmetric~\cite{Irving-Kirkwood}.
Let $\tau_{\rm micro}$ denote the correlation time of the force acting on the top plate at $z=L_{z}$.
Throughout this paper, for any physical quantity $A(\bm{r},t)$, we define the finite time-averaged quantity as
\begin{equation}
 \bar{A}(\bm{r}) \equiv \frac{1}{\tau}\int_{0}^{\tau}\mathrm{d}t\; A(\bm{r},t),
\end{equation}
where $\tau$ is taken to satisfy $\tau \gg \tau_{\mathrm{micro}}$.
We define the $zx$- and $zz$-component of the time-averaged stress tensor per unit area on the top plate as
\begin{equation}
 \bar{\sigma}_{\mathrm{S}} = \frac{1}{L_{x}L_{y}}\int_{0}^{L_{x}}\mathrm{d}x\int_{0}^{L_{y}}\mathrm{d}y\; \bar{\sigma}_{zx}(x,y,L_{z}),
  \label{eq:def_sigmaast_S}
\end{equation}
and
\begin{equation}
 \bar{\sigma}_{\mathrm{N}} = \frac{1}{L_{x}L_{y}}\int_{0}^{L_{x}}\mathrm{d}x\int_{0}^{L_{y}}\mathrm{d}y\; \bar{\sigma}_{zz}(x,y,L_{z}),
  \label{eq:def_sigmaast_N}
\end{equation}
respectively.
We study the fluctuations of $\bar{\sigma}_{\mathrm{S}}$ and $\bar{\sigma}_{\mathrm{N}}$ in the equilibrium system at temperature $T$.
For later convenience, we define the deviation of $\bar{\sigma}_{ab}$ from its equilibrium value as
\begin{equation}
 \bar{s}_{ab}(\bm{r}) = \bar{\sigma}_{ab}(\bm{r}) + p_{\mathrm{th}} \delta_{ab},
  \label{eq:devi_stress}
\end{equation}
where $p_{\mathrm{th}}$ is the thermodynamic pressure satisfying $\aves{\bar{\sigma}_{ab}(\bm{r})}_{\mathrm{eq}}=-p_{\mathrm{th}}\delta_{ab}$.

\subsection{Assumptions} \label{subsec:assump}

We assume that (i) the time-averaged fluctuating stress is balanced in equilibrium at each position, which is expressed as
\begin{equation}
 \nabla \cdot \bar{\bm{\sigma}}(\bm{r})=\bm{0}.
  \label{eq:stress_balance}
\end{equation}
We also assume that (ii) $\bar{\bm{\sigma}}$ obeys a Gaussian distribution on the restricted configuration space given by (\ref{eq:stress_balance}).
Because $\bar{\bm{\sigma}}$ is obtained by integrating the microscopic stress over a suitable region, it is reasonable, by considering the central limit theorem, to assume (ii).
Currently, it is difficult to derive assumption (ii) from a microscopic mechanical system.
Note that, in our previous paper~\cite{Itami-Sasa2015}, we derived Stokes' law by using assumptions (i) and (ii).

\subsection{Probability density of the time-averaged stress fields}

In this section, we derive the probability density of the time-averaged stress fields in the equilibrium viscous fluids.
According to the Green--Kubo formula~\cite{Green1954}, the shear viscosity $\eta$ and the volume viscosity $\zeta$ are related to the stress fluctuations in the bulk of the fluid as
\begin{eqnarray}
 \ave{ \bar{\sigma}_{xy}(\bm{r}) \bar{\sigma}_{xy}(\bm{r}')}_{\mathrm{eq}} = \frac{2k_{\mathrm{B}}T \eta}{\tau} \delta(\bm{r}-\bm{r}'),
 \label{eq:GK_s}
 \\[3pt]
 \ave{ \frac{\sum_{a}\bar{s}_{aa}(\bm{r})}{3} \frac{\sum_{b}\bar{s}_{bb}(\bm{r}')}{3}}_{\mathrm{eq}} = \frac{2k_{\mathrm{B}}T \zeta}{\tau} \delta(\bm{r}-\bm{r}'),
 \label{eq:GK_v}
\end{eqnarray}
where $\aves{\; \cdot\; }_{\mathrm{eq}}$ denotes a canonical ensemble average at temperature $T$, $k_{\mathrm{B}}$ is the Boltzmann constant, and $\tau$ is assumed to be much larger than the correlation time of the stress tensor.
Using the Green--Kubo formula in (\ref{eq:GK_s}) and (\ref{eq:GK_v}), the symmetric property of the stress tensor $\bar{\sigma}_{ab}=\bar{\sigma}_{ba}$, and the isotropic property, we obtain the statistical property of $\bar{\bm{\sigma}}$:
\begin{eqnarray}
 \aves{\bar{s}_{ab}(\bm{r})\bar{s}_{a'b'}(\bm{r}')}_{\mathrm{eq}} = \frac{2k_{\mathrm{B}}T}{\tau} \Delta_{aba'b'} \delta (\bm{r}-\bm{r}')
 \label{eq:s_cov}
\end{eqnarray}
with
\begin{equation}
 \Delta_{aba'b'} = \eta\delta_{aa'}\delta_{bb'} + \eta\delta_{ab'}\delta_{ba'} + \left( \zeta-\frac{2}{3}\eta \right)\delta_{ab}\delta_{a'b'}.
\end{equation}
We express  $\Delta_{aba'b'} =\Delta_{ij}$ with $g(ab)=i$ and $g(a'b')=j$, where $g(xx)=1$, $g(yy)=2$, $g(zz)=3$, $g(xy)=g(yx)=4$, $g(yz)=g(zy)=5$, and $g(zx)=g(xz)=6$. 
That is, $\Delta$ is interpreted as a $6 \times 6$ matrix.
The inverse of $\Delta$ is given by
\begin{equation}
\Delta^{-1}= \left(   
 \begin{array}{cccccc}
  \hspace*{3.3mm}\frac{1}{3\eta}+\frac{1}{9\zeta}\; & -\frac{1}{6\eta}+\frac{1}{9\zeta}\; & -\frac{1}{6\eta}+\frac{1}{9\zeta}\; & \; 0\; & \; 0\; & \; 0\; \\[1.5mm]
  -\frac{1}{6\eta}+\frac{1}{9\zeta}\; & \hspace*{3.3mm} \frac{1}{3\eta}+\frac{1}{9\zeta}\; & -\frac{1}{6\eta}+\frac{1}{9\zeta}\; & 0 & 0 & 0 \\[1.5mm]
  -\frac{1}{6\eta}+\frac{1}{9\zeta}\; & -\frac{1}{6\eta}+\frac{1}{9\zeta}\; & \hspace*{3.3mm}\frac{1}{3\eta}+\frac{1}{9\zeta}\; & 0 & 0 & 0 \\[1.5mm]
  0 & 0 & 0 & \frac{1}{\eta} & 0 & 0 \\[1.5mm]
  0 & 0 & 0 & 0 & \frac{1}{\eta} & 0 \\[1.5mm]
  0 & 0 & 0 & 0 & 0 & \frac{1}{\eta}
 \end{array}
 \right).
\end{equation}
We identify $\Delta^{-1}_{aba'b'}$ with $(\Delta^{-1})_{ij}$ by $i=g(ab)$ and $j=g(a'b')$.
From assumptions (i) and (ii), the probability density of the time-averaged stress tensor in equilibrium is given by
\begin{eqnarray}
 \mathcal{P}\left( \{ \bar{\sigma}_{ab}(\bm{r})\} \right) = C \exp \left[ - \tau \mathcal{I}\left( \{ \bar{\sigma}_{ab}(\bm{r})\} \right) \right]  \prod_{\bm{r}}\delta \Big( \nabla \cdot \bar{\bm{\sigma}}(\bm{r})\Big) 
  \label{eq:pro_all_stress}
\end{eqnarray}
with 
\begin{eqnarray}
 \mathcal{I}\left( \{ \bar{\sigma}_{ab}(\bm{r})\} \right) = \frac{1}{4k_{\mathrm{B}}T} \int_{0}^{L_{x}}\hspace{-2mm}\mathrm{d}x \int_{0}^{L_{y}}\hspace{-2mm}\mathrm{d}y \int_{0}^{L_{z}}\hspace{-2mm}\mathrm{d}z \sum_{i,j} \bar{s}_{ab}(\bm{r})\Delta^{-1}_{aba'b'}\bar{s}_{a'b'}(\bm{r}) ,
 \label{eq:LD_sigma}
\end{eqnarray}
where $C$ is a normalization constant.
Note that we can explicitly write the integrand on the right-hand side of (\ref{eq:LD_sigma}) as
\begin{eqnarray}
 \fl \quad \sum_{i,j} \bar{s}_{ab}\Delta^{-1}_{aba'b'}\bar{s}_{a'b'} &= \frac{(\bar{\sigma}_{xx}-\bar{\sigma}_{yy})^{2}+(\bar{\sigma}_{yy}-\bar{\sigma}_{zz})^{2}+(\bar{\sigma}_{zz}-\bar{\sigma}_{xx})^{2}}{6\eta}
 \nonumber \\
 & \qquad + \frac{(\bar{\sigma}_{xy})^{2}+(\bar{\sigma}_{yz})^{2}+(\bar{\sigma}_{zx})^{2}}{\eta} + \frac{1}{\zeta}\left( \frac{\bar{\sigma}_{xx}+\bar{\sigma}_{yy}+\bar{\sigma}_{zz}}{3}+p_{\mathrm{th}}\right)^{2}.
 \label{eq:sum_sigma}
\end{eqnarray}

\section{Results} \label{sec:results}

\subsection{Time-averaged shear stress fluctuations} \label{subsec:result_S}

In this section, we calculate $\aves{( \bar{\sigma}_{\mathrm{S}} )^2 }_{\mathrm{eq}}$.
From $\mathcal{P}\left( \{ \bar{\sigma}_{ab}(\bm{r})\} \right)$, we formally obtain the probability density of $\bar{\sigma}_{\mathrm{S}}$:
\begin{eqnarray}
 \fl \quad P(\bar{\sigma}_{\mathrm{S}}) = \int \prod_{i}\mathcal{D} \bar{\sigma}_{ab}\; \mathcal{P}\left( \{ \bar{\sigma}_{ab}(\bm{r})\} \right) \delta \left( \bar{\sigma}_{\mathrm{S}} - \frac{1}{L_{x}L_{y}}\int_{0}^{L_{x}}\hspace{-2mm}\mathrm{d}x\int_{0}^{L_{y}}\hspace{-2mm}\mathrm{d}y\; \bar{\sigma}_{zx}(x,y,L_{z})\right).
  \label{eq:pro_sigmaast_S}
\end{eqnarray}
In the asymptotic regime $\tau\gg\tau_{\rm micro}$, the functional integral may be accurately evaluated by the saddle-point method.
By introducing the Lagrange multiplier field $\bm{\lambda}(\bm{r})= (\lambda_{x}(\bm{r}), \lambda_{y}(\bm{r}),\lambda_{z}(\bm{r}))$ to take constraint (\ref{eq:stress_balance}) into account, we obtain
\begin{equation}
 P(\bar{\sigma}_{\mathrm{S}}) = C_{\ast} \exp \left[ -\tau I(\bar{\sigma}_{\mathrm{S}}) \right] 
\label{eq:pro_sur_stress}
\end{equation}
with
\begin{equation}
 I(\bar{\sigma}_{\mathrm{S}}) = \min_{\{ \bar{\sigma}_{ab}(\bm{r}): (\ref{eq:def_sigmaast_S})\} }\; \frac{1}{4k_{\mathrm{B}}T}\int_{0}^{L_{x}}\mathrm{d}x \int_{0}^{L_{y}} \mathrm{d}y \int_{0}^{L_{z}} \mathrm{d}z \; \mathcal{L}\left( \{ \bar{\sigma}_{ab}(\bm{r})\} \right) ,
  \label{eq:contraction_stress_S}
\end{equation}
and
\begin{eqnarray}
 \mathcal{L}\left( \{ \bar{\sigma}_{ab}(\bm{r})\} \right) = \sum_{i,j}\bar{s}_{ab}(\bm{r})\Delta^{-1}_{aba'b'}\bar{s}_{a'b'}(\bm{r})  + \sum_{a,b}2 \lambda_{a}(\bm{r}) \partial_{b}\bar{\sigma}_{ab}(\bm{r}),
 \label{eq:Lagrangian_S}
\end{eqnarray}
where $C_{\ast}$ is a normalization constant, and $\min_{\{ \bar{\sigma}_{ab}(\bm{r}): (\ref{eq:def_sigmaast_S})\} }(\cdots )$ means that we minimize $(\cdots )$ under the condition (\ref{eq:def_sigmaast_S}).
Note that the saddle-point method for the large deviation function has been rigorously verified under certain conditions. 
This is called the contraction principle~\cite{Touchette} in probability theory.

We next solve the variational problem in (\ref{eq:contraction_stress_S}).
Because the boundary condition for $\bar{\sigma}_{ab}(\bm{r})$ is only (\ref{eq:def_sigmaast_S}), $\bm{\lambda}$ must satisfy
\begin{equation}
 \sum_{a} \int_{0}^{L_{x}}\mathrm{d}x\int_{0}^{L_{y}}\mathrm{d}y\int_{0}^{L_{z}}\mathrm{d}z\; \partial_{z}\left[ \lambda_{a}(\bm{r})\delta\bar{\sigma}_{az}(\bm{r})\right] = 0,
\end{equation}
for any $\delta \bar{\sigma}_{az}(\bm{r})$, where $\delta \bar{\sigma}_{az}(\bm{r})$ is a variation of $\bar{\sigma}_{az}(\bm{r})$.
We thus obtain the following conditions:
\begin{equation}
 \bm{\lambda}(x,y,0) = \bm{0},\quad \lambda_{x}(x,y,L_{z}) = V_{0},\quad \lambda_{y}(x,y,L_{z}) = \lambda_{z}(x,y,L_{z}) = 0,
  \label{eq:bcond_S}
\end{equation}
where $V_{0}$ is a constant.
These conditions are called the natural boundary conditions.
Note that $V_{0}$ need not be zero due to (\ref{eq:def_sigmaast_S}).

Under the natural boundary conditions (\ref{eq:bcond_S}), the surface contribution of the variation vanishes, and we obtain the Euler--Lagrange equations
\begin{equation}
 \frac{\partial \mathcal{L}}{\partial \bar{\sigma}_{ab}} - \sum_{c}\partial_{c} \left( \frac{\partial \mathcal{L}}{\partial \left( \partial_{c}\bar{\sigma}_{ab}\right)}\right) = 0,
  \label{eq:E-L_eq}
\end{equation}
which are equivalent to
\begin{equation}
 \bar{\sigma}_{ab}(\bm{r}) = \left[ -p_{\mathrm{th}}+\left(\zeta -\frac{2}{3}\eta\right) \nabla\cdot\bm{\lambda}(\bm{r})\right] \delta_{ab} + \eta \left( \partial_{a}\lambda_{b}(\bm{r})+\partial_{b}\lambda_{a}(\bm{r})\right).
  \label{eq:vari_stress_pre_S}
\end{equation}
The Helmholtz decomposition of $\bm{\lambda}(\bm{r})$ leads to
\begin{equation}
 \bm{\lambda}(\bm{r}) = -\nabla \phi(\bm{r}) + \bm{u}(\bm{r})
\end{equation}
with
\begin{equation}
 \nabla\cdot\bm{u}(\bm{r}) = 0.
  \label{eq:con_incom}
\end{equation}
Then, (\ref{eq:vari_stress_pre_S}) can be rewritten as
\begin{equation}
 \bar{\sigma}_{ab}=-\left[ p_{\mathrm{th}}+\left(\zeta -\frac{2}{3}\eta\right) \nabla^2 \phi\right] \delta_{ab} - 2\eta\partial_{a}\partial_{b}\phi + \eta \left( \partial_{a}u_{b}+\partial_{b}u_{a}\right).
  \label{eq:vari_stress_S}
\end{equation}
Substituting this expression into (\ref{eq:stress_balance}), we obtain
\begin{equation}
 \eta \nabla^{2} \bm{u}(\bm{r}) = \nabla p(\bm{r})
  \label{eq:Stokes_eq}
\end{equation}
with
\begin{equation}
 p(\bm{r}) = p_{\mathrm{th}} + \left(\zeta +\frac{4}{3}\eta\right) \nabla^2 \phi(\bm{r}).
  \label{eq:eff_pres}
\end{equation}
Thus, we may interpret $\bm{u}(\bm{r})$ as the virtual velocity fields of the fluid generated by the most probable stress fields for a given $\bar{\sigma}_{\mathrm{S}}$.
We may also interpret $p(\bm{r})$ as the virtual pressure field.
By solving (\ref{eq:con_incom}), (\ref{eq:vari_stress_S}), (\ref{eq:Stokes_eq}), and (\ref{eq:eff_pres}) with the boundary conditions (\ref{eq:def_sigmaast_S}) and (\ref{eq:bcond_S}), we can obtain $I(\bar{\sigma}_{\ast})$.
Note that the boundary conditions in (\ref{eq:bcond_S}) are not related to boundary conditions for the real velocity fields of the fluid.

By referring to the solution of the Stokes equations with the boundary conditions (\ref{eq:def_sigmaast_S}) and (\ref{eq:bcond_S}), we obtain
\begin{equation}
 u_{x} (\bm{r}) = \frac{\bar{\sigma}_{\mathrm{S}}z}{\eta}, \quad u_{y}(\bm{r}) = u_{z}(\bm{r}) = 0, \quad \nabla \phi (\bm{r})= \bm{0}.
\end{equation}  
These results with (\ref{eq:vari_stress_S}) lead to
\begin{equation}
 \fl \quad \bar{\sigma}_{xx}(\bm{r}) = \bar{\sigma}_{yy}(\bm{r}) = \bar{\sigma}_{zz}(\bm{r}) = -p_{\mathrm{th}} ,\quad \bar{\sigma}_{xy}(\bm{r}) = \bar{\sigma}_{yz}(\bm{r}) = 0, \quad \bar{\sigma}_{zx}(\bm{r}) = \bar{\sigma}_{\mathrm{S}}.
  \label{eq:mpstress_S}
\end{equation}
Thus, from (\ref{eq:contraction_stress_S}), we obtain
\begin{eqnarray}
 I(\bar{\sigma}_{\mathrm{S}}) = \frac{L_{x}L_{y}L_{z}}{4k_{\mathrm{B}}T\eta} (\bar{\sigma}_{\mathrm{S}})^2,
\end{eqnarray}
which leads to
\begin{equation}
 \ave{(\bar{\sigma}_{\mathrm{S}})^2}_{\mathrm{eq}} = \frac{2k_{\mathrm{B}}T\eta}{L_{x}L_{y}L_{z} \tau} .
  \label{eq:vari_sigma_S}
\end{equation}

Because $\bar{\sigma}_{\mathrm{S}}$ is a time- and surface-averaged quantity, we naively expect the central limit theorem:
\begin{equation}
 \ave{(\bar{\sigma}_{\mathrm{S}})^2}_{\mathrm{eq}} \propto \frac{1}{L_{x}L_{y}\tau},
  \label{eq:naive_CLT_S}
\end{equation}
where $\propto$ implies that we only focus on the $L_{x}L_{y}$ and $\tau$ dependencies of the function.
With regard to these dependencies, the result in (\ref{eq:vari_sigma_S}) is consistent with (\ref{eq:naive_CLT_S}).
In addition to this standard behaviour, (\ref{eq:vari_sigma_S}) implies that the fluctuations of $\bar{\sigma}_{\mathrm{S}}$ are inversely proportional to $L_{z}$, which may originate from long-range stress correlations via the virtual Couette flow generated by the most probable stress fields (\ref{eq:mpstress_S}) for a given $\bar{\sigma}_{\mathrm{S}}$.
Such $L_{z}$ dependence is singular because the fluctuations vanish in the limit $L_{z}\to \infty$.
Furthermore, we may evaluate the vertical system size $L_{z}$ by only measuring the time- and surface-averaged stress fluctuation on the top plate in equilibrium.
We stress that assumption (i) expressed as (\ref{eq:stress_balance}) is essential to the singular behaviour.
Note that the singular behaviour was observed in molecular dynamics simulations~\cite{Petravic-Harrowell}.

\subsection{Time-averaged normal stress fluctuations} \label{subsec:result_N}

In this section, we calculate $\aves{( \bar{\sigma}_{\mathrm{N}}+p_{\mathrm{th}} )^2 }_{\mathrm{eq}}$.
Using the same method as in section \ref{subsec:result_S}, we obtain the large deviation function of $\bar{\sigma}_{\mathrm{N}}$:
\begin{equation}
 I(\bar{\sigma}_{\mathrm{N}}) = \min_{\{ \bar{\sigma}_{ab}(\bm{r}) : (\ref{eq:def_sigmaast_N})\} }\; \frac{1}{4k_{\mathrm{B}}T}\int_{0}^{L_{x}}\mathrm{d}x \int_{0}^{L_{y}} \mathrm{d}y \int_{0}^{L_{z}} \mathrm{d}z \; \mathcal{L}\left( \{ \bar{\sigma}_{ab}(\bm{r})\} \right) ,
  \label{eq:contraction_stress_N}
\end{equation}
with (\ref{eq:Lagrangian_S}).

Next, we solve the variational problem in (\ref{eq:contraction_stress_N}).
To satisfy
\begin{equation}
 \sum_{a} \int_{0}^{L_{x}}\mathrm{d}x\int_{0}^{L_{y}}\mathrm{d}y\int_{0}^{L_{z}}\mathrm{d}z\; \partial_{z}\left[ \lambda_{a}(\bm{r})\delta\bar{\sigma}_{az}(\bm{r})\right] = 0,
\end{equation}
for any $\delta\bar{\sigma}_{az}(\bm{r})$, the natural boundary conditions are given by
\begin{equation}
 \bm{\lambda}(x,y,0) = \bm{0},\quad \lambda_{x}(x,y,L_{z}) = \lambda_{y}(x,y,L_{z}) = 0,\quad \lambda_{z}(x,y,L_{z}) = V_{0}, 
  \label{eq:bcond_N}
\end{equation}
where $V_{0}$ is a constant.
Then, the Euler--Lagrange equations are given by
\begin{equation}
 \eta \nabla^{2} \bm{\lambda} + \left( \zeta + \frac{\eta}{3}\right)\nabla \left( \nabla\cdot\bm{\lambda}\right) = 0
  \label{eq:Stokes_eq_N}
\end{equation}
with
\begin{equation}
 \bar{\sigma}_{ab}(\bm{r}) = \left[ -p_{\mathrm{th}}+\left(\zeta -\frac{2}{3}\eta\right) \nabla\cdot\bm{\lambda}(\bm{r})\right] \delta_{ab} + \eta \left( \partial_{a}\lambda_{b}(\bm{r})+\partial_{b}\lambda_{a}(\bm{r})\right).
  \label{eq:vari_stress_N}
\end{equation}
Recalling the boundary conditions (\ref{eq:def_sigmaast_N}) and (\ref{eq:bcond_N}), we have
\begin{eqnarray}
 \lambda_{x}(\bm{r}) &= \lambda_{y}(\bm{r}) = 0,\quad \lambda_{z}(\bm{r})= \frac{\bar{\sigma}_{\mathrm{N}}+p_{\mathrm{th}}}{\zeta+\frac{4}{3}\eta}z,
\end{eqnarray}
and
\begin{equation}
\eqalign{
 \bar{\sigma}_{xx}(\bm{r}) &= \bar{\sigma}_{yy}(\bm{r}) = -p_{\mathrm{th}}+\frac{\zeta-\frac{2}{3}\eta}{\zeta+\frac{4}{3}\eta}\left( \bar{\sigma}_{\mathrm{N}}+p_{\mathrm{th}}\right),
 \cr
 \bar{\sigma}_{zz}(\bm{r}) &= \bar{\sigma}_{\mathrm{N}}, \quad \bar{\sigma}_{xy}(\bm{r}) = \bar{\sigma}_{yz}(\bm{r}) = \bar{\sigma}_{zx}(\bm{r}) = 0.
}
\label{eq:mpstress_N}
\end{equation}
Thus, from (\ref{eq:contraction_stress_N}), we obtain
\begin{eqnarray}
 I(\bar{\sigma}_{\mathrm{N}}) &= \frac{L_{x}L_{y}L_{z}}{4k_{\mathrm{B}}T\left( \zeta+\frac{4}{3}\eta\right)} \left( \bar{\sigma}_{\mathrm{N}}+p_{\mathrm{th}}\right)^2,
\end{eqnarray}
which leads to
\begin{equation}
 \ave{(\bar{\sigma}_{\mathrm{N}}+p_{\mathrm{th}})^2}_{\mathrm{eq}} = \frac{2k_{\mathrm{B}}T\left( \zeta+\frac{4}{3}\eta\right)}{L_{x}L_{y}L_{z}\tau}.
  \label{eq:vari_sigma_N}
\end{equation}
The time-averaged normal stress fluctuations on the surface also depend singularly on $L_{z}$.
This result is similar to that reported in Ref.~\cite{MonahanETAL}.
Because $\aves{(\bar{\sigma}_{\mathrm{N}}+p_{\mathrm{th}})^2}_{\mathrm{eq}}$ depends on $\zeta$, it is necessary for our calculation to introduce $\zeta$.
Note that we obtain the same result as (\ref{eq:vari_sigma_S}) when we use the method of \cite{Itami-Sasa2015}, where the effect of the volume viscosity $\zeta$ is ignored.
Unlike the time-averaged normal stress fluctuations in the bulk (\ref{eq:GK_v}), the shear viscosity $\eta$ is relevant for describing the time-averaged normal stress fluctuations on the surface.

\section{Concluding Remarks}

We have provided the new method for calculating time-averaged stress fluctuations on surfaces in a viscous incompressible fluid at equilibrium by using the contraction principle.
We have shown that the time-averaged shear and normal stress fluctuations on the plates are inversely proportional to the distance between the plates.

Before ending this paper, we wil make some remarks.
The most important future study will be to observe (\ref{eq:vari_sigma_S}) and (\ref{eq:vari_sigma_N}) experimentally.
The singular $L_{z}$ dependence and the anisotropic nature characterized by the volume viscosity are so qualitatively new that they warrant verification with experiments.
Theoretically, our method is still at a primitive level.
For example, the generalization to more complicated geometries should be formulated.
Furthermore, an understanding of the connection between our approach and the method using fluctuating hydrodynamics in Ref.~\cite{MonahanETAL} has not been elucidated.
We believe that the theory developed in this paper may be a first step towards an understanding of the singular behaviour of stress fluctuations in fluids.

\ack
The authors thank H. Nakano for useful discussions. 
The present study was supported by JSPS KAKENHI Grant Numbers JP17K14355, JP17H01148.

\end{document}